\documentclass[3p,draft]{elsarticle}
\usepackage[T1]{fontenc}
\usepackage{graphicx, algorithm, algorithmic}
\usepackage[utf8]{inputenc}
\usepackage{tikz}
\usetikzlibrary{shapes.geometric,shapes.misc}
\usetikzlibrary{calc,shapes,arrows,arrows.meta,shapes.misc,decorations.pathreplacing}
\usetikzlibrary{positioning,calc}
\usepackage{tikz-qtree}
\usepackage{subcaption}
\usepackage{lipsum}
\usepackage{listings}
\usepackage{adjustbox}
\tikzset{align at top/.style={baseline=(current bounding box.north)}}
\tikzstyle{csucs}=[circle,draw,minimum size=0.6cm,inner sep=0.5pt]
\tikzset{>={triangle 45}}

\usepackage{amsmath,amssymb}

\newtheorem{proposition}{Proposition}
\newtheorem{definition}{Definition}
\newtheorem{theorem}{Theorem}
\newtheorem{lemma}{Lemma}
\newtheorem{example}{Example}
\newproof{proof}{Proof}

\begin{document}

\title{Maintaning maximal matching with lookahead\tnoteref{thanksfn}}
\author[kitti]{Kitti~Gelle}
\author[szabivan]{Szabolcs~Iv\'an}
\address[kitti,szabivan]{University of Szeged, Hungary}
\tnotetext[thanksfn]{Kitti Gelle was supported by the \'UNKP-17-3-I-SZTE-18 New National Excellence Program of the Ministry of Human Capacities. Szabolcs Iv\'an was supported by NKFI grant number K108448.} 

\def\cP{\ensuremath{\mathcal{P}}}
\def\dfs{\ensuremath{\textsc{dfs}}}
\def\nul{{\textsc{null}}}
\def\mate{{\textsc{mate}}}
\def\insert{{\textsc{insert}}}
\def\remove{{\textsc{delete}}}

\begin{abstract}
	In this paper we study the problem of fully dynamic maximal matching with lookahead. In a fully dynamic $n$-vertex graph setting, we have to handle updates (insertions and removals of edges), and answer
	queries regarding the current graph, preferably with a better time bound than that when running the trivial deterministic algorithm with worst-case time of $O(m)$ (where $m$ is the all-time maximum number of the edges) and recompute the matching from scratch each time a query arrives.
	We show that a maximal matching can be maintained in an (undirected) general graph
	with a deterministic amortized update cost of $O(\log m)$,
	provided that a lookahead of length $m$ is available, i.e. we can ``take a peek'' at the next $m$ update operations
	in advance.
\end{abstract}

\maketitle
\section{Introduction and notation}
Graph algorithms are fundamental in computer science. In most cases, graphs have been studied as static objects, however, in many real life examples (e.g. social networks, AI) they are changing in size. In the last few decades, there has been a growing interest in developing algorithms and data structures for such \emph{dynamic graphs}. In this setting, graphs are subject to \emph{updates} -- in our case, additions and removals of an edge at a time.
The aim of a so-called fully dynamic algorithm (here ``fully'' means that both addition and removal are
supported) is to maintain the result of the algorithm after each and every update of the graph,
in a time bound significantly better than recomputing it from scratch each time.

In~\cite{lookahead}, a systematic investigation of dynamic graph problems in the
presence of a so-called \emph{lookahead} was initiated: although the stream of update operations
can be arbitrarily large and possibly builds up during the computation time, in actual real-time
systems it is indeed possible to have some form of \emph{lookahead} available.
That is, the algorithm is provided with some prefix of the update sequence of some length
(for example, in~\cite{lookahead} an assembly planning problem is studied in which the
algorithm can access the prefix of the sequence of future operations to be handled of length $\Theta(\sqrt{m/n}\log n)$), where $m$ and $n$ are the number of edges and nodes, respectively.
Similarly to the results of~\cite{lookahead} (where the authors devised dynamic algorithms using lookahead
for the problems of strongly connectedness and transitive closure), we will execute the tasks in batches:
by looking ahead at $O(m)$ future update operations, we treat them as a single batch, preprocess
our current graph based on the information we get from the complete batch, then we run
all the updates, one at a time, on the appropriately preprocessed graph.
This way, we achieve an amortized update cost of $O(\log m)$ for maintaining a maximal matching.

We view a graph $G$ as a set (or list) of edges, with $|G|$ standing for its cardinality.
This way notions like $G\cup H$ for two graphs $G$ and $H$ (sharing the common set $V(G)=V(H)$ of vertices)
are well-defined.

In the studied problem,
a \emph{matching} of a(n undirected) graph $G$ is a subset $M\subseteq G$ of edges having pairwise
disjoint sets of endpoints. A matching $M$ is \emph{maximal} if there is no matching $M'\supsetneq M$ of $G$.
Given a matching $M$, for each vertex $v$ of $G$ let $\textsc{mate}(v)$ denote the unique vertex $u$ such that
$(u,v)\in M$ if such a vertex exists, otherwise $\textsc{mate}(v)=\nul$.

In the fully dynamic version of the maximal matching problem, the update operations are edge additions
$+(u,v)$, edge deletions $-(u,v)$ and the queries have the form $\textsc{mate}(u)$.

So a fully dynamic algorithm for maximal matching problem supports the following operations on an undirected graph $G=(V,E)$:
\begin{itemize}
	\item \textbf{\insert}$(u,v)$: inserts an edge between $u$ and $v$
	\item \textbf{\remove}$(u,v)$: deletes an edge between $u$ and $v$
	\item \textbf{\textsc{mate}}$(u)$: answers $v$ if $(u,v)\in M$, where $M$ is the current maximal matching of $G$ and $\nul$ otherwise
\end{itemize} 

{\bf Related work.}
There is an interest in computing a maximum (i.e. maximum cardinality) or maximal (i.e. non-expandable) matching
in the fully dynamic setting. There is no ``best-so-far'' algorithm, since the settings differ:
Baswana, Gupta and Sen \cite{baswana} presented a randomized algorithm for maximal matching,
having an $O(\log n)$ expected amortized time
per update. Based on this algorithm Solomon \cite{solomon} gave a randomized algorithm with constant  amortized update time. (Note that algorithms for maximal matching automatically provide $2$-approximations for
maximum matching and also vertex cover.) For the deterministic variant, Ivkovi\`c and Lloyd \cite{ivkovic} defined
an algorithm with an $O((n+m)^{0.7072})$ amortized update time, which was improved to
an amortized $O(\sqrt{m})$ update cost by Neiman and Solomon~\cite{neiman}.
For maximum matching, Onak and Rubinfeld \cite{onak} developed a randomized algorithm that achieves a $c$-approximation
for some constant $c$, with an $O(\log^2 n)$ expected amortized update time.
To maintain an exact maximum cardinality matching, Micali and Vazirani \cite{micali} gave an algorithm
with a worst-case update time of $O(\sqrt{n}\cdot m)$. Allowing randomization, an update cost of $O(n^{1.495})$
is achievable due to Sankowski~\cite{sankowski}.

We are not aware of any results on allowing lookahead for any of the matching problems,
but the notion has been applied to several problems in this field: following the seminal work of Khanna, Motwani and Wilson~\cite{lookahead},
where lookahead was investigated for the problems of maintaining the transitive closure and the 
strongly connectedness of a directed graph, Sankowski and Mucha~\cite{sankowskimucha} also considered the transitive closure
with lookahead via the dynamic matrix inverse problem,
devising a randomized algorithm, and Kavitha~\cite{kavitha} studied the dynamic matrix rank problem.

\section{Maximal matching with lookahead}

In this section we present an algorithm that maintains a maximal matching in a dynamic graph $G$
with constant query and $O(\log m)$ update time (note that $O(\log m)$ is also $O(\log n)$ as $m=O(n^2)$),
provided that a \emph{lookahead} of length $m$ is available in the sequence of (update and query) operations.
This is an improvement over the currently best-known deterministic algorithm~\cite{neiman} that has an update cost of $O(\sqrt{m})$
without lookahead.

The following is clear:
\begin{proposition}
	Suppose $G$ is a graph in which $M$ is a maximal matching.
	Then a maximal matching in the graph $G+(u,v)$ is
	\begin{itemize}
		\item $M\cup\{(u,v)\}$, if $\mate(u)=\mate(v)=\nul$,
		\item $M$, otherwise.
	\end{itemize}
\end{proposition}
This proposition gives the base algorithm $\textsc{greedy}$ for computing a maximal matching in a graph:
\begin{lstlisting}[language=Java,mathescape=true]
Let $M$ be an empty list of edges;
for( $(u,v)\in G$ ) {
  if( $\mate(u)==\nul$ and $\mate(v)==\nul$){
    $\mate(u):=v$; $\mate(v):=u$;
    insert $(u,v)$ to $M$;
  }
}
return $M$;
\end{lstlisting}
Note that if one initializes the $\mate$ array in the above code so that it contains some non-$\nul$ entries,
then the result of the algorithm represents a maximal matching within the subgraph of $G$ spanned
by the vertices having $\nul$ $\mate$s initially. Also, with $M$ represented by a linked list,
the above algorithm runs in $O(m)$ total time using no lookahead.
Hence, by calling this algorithm on each update operation (after inserting or removing the edge in question),
we get a dynamic graph algorithm with no lookahead (hence it uses a lookahead of at most $m$ operations),
a constant query cost (as it stores the $\mate$ array explicitly) and an $O(m)$ update cost.
Using this algorithm $A_1$, we build up a sequence $A_k$ of algorithms, each having a smaller update cost
than the previous ones. (In a practical implementation there would be a single algorithm $A$ taking
$k$ as a parameter along with the graph $G$ and the update sequence, but for proving the time complexity
it is more convenient to denote the algorithms in question by $A_1$, $A_2$, and so on.)

In our algorithm descriptions the input is the current graph $G$ (which is $\emptyset$ the first time we
start running the program) and a sequence $(q_1,\ldots,q_t)$ of operations. Of course as the sequence can be
arbitrarily long, we do not require an explicit representation, just the access of the first $m$ elements
(that is, we have a lookahead of length $m$).

To formalize our main lemma in a more concise way, we first
define the invariant property, which we call \emph{$h(m)$-ensuring},
of these algorithms:
\begin{definition}
	We say that an algorithm $A$ is an \emph{$h(m)$-ensuring algorithm for maximal matching}, if $A$ is a fully dynamic algorithm maintaining a maximal matching
	in a graph such that if it gets as input a graph $G$, as an edge list, having
	$m_0$ edges initially, and a (possibly infinite) stream $(q_1,q_2,\ldots,q_t)$
	of updates with $t\geq m_0$, then $A$ can process these queries with an
	amortized update cost of $h(m)$ using a lookahead of length $m$,
	such that between handling of these updates,
	queries of the form $\mate(u)$, asking for the mate of vertex $u$ in the
	current maximal matching, can be answered in a constant time.
\end{definition}
In the definition above, $m$ stands for the maximum number of edges in $G$
during its life cycle, formally, $m:=\max\{|Gq_1q_2\ldots q_i|:0\leq i\leq t\}$.

As an example, the following algorithm $A_0$
that runs $\textsc{Greedy}$ after each update,
is a $c\cdot m$-ensuring algorithm for maximal matching, for some universal constant
$c$:
\begin{enumerate}
	\item Initialize a global array $\mate$ of vertices, set $\mate(u):=\nul$
	for each vertex $u$.
	\item Upon receiving an update sequence $(q_1,\ldots,q_t)$, the algorithm
	  does the following:
	\begin{enumerate}
		\item Let $M$ be an empty list of edges.
		\item For processing $q_i$, we
		\begin{enumerate}
			\item first modify $G$ accordingly, $G:=G\cdot q_i$,
			\item then we iterate through the current matching $M$ and
			  set $\mate(u)=\mate(v)=\nul$ for each $(u,v)\in M$, emptying $M$
			  during the process,
			\item we set $M:=\textsc{Greedy}(G,\mate)$.
		\end{enumerate}
		\item Having processed $q_i$, we now can answer queries of the form
		  $\mate(u)$ in a constant time, by accessing the global array $\mate$.
	\end{enumerate}
\end{enumerate}
Step $1$ has a setup cost of $O(n)$. When we receive the update sequence, the
local initialization of $M$ takes a constant time. Note that for processing $q_i$
we do not use any lookahead which is fine. Modifying the current graph $G$
in Step 2.b.i.
takes $O(m)$ time, since adding/removing an entry to a list of unique entries
takes a time proportional to the size of the list, which is by definition of $m$,
at most $m$ at any given time point $i$. Then, as the matching $M$ is also a
list of at most $m$ edges, iterating through it takes $O(m)$ iterations,
setting the $\mate$ array for a constant time each, so Step 2.b.ii. also
takes $O(m)$ time. Finally, Step 2.b.iii. also takes $O(m)$ time, and after
that, we clearly have a maximal matching for $G_i:=Gq_1\ldots q_i$, stored
in the $\mate$ array. The total cost for handling a single update is thus
$c\cdot m$ for some universal constant $c$.

Note after in each step we erase our ``local'' matching $M$ from the $\mate$
array for a total cost of $O(m)$ since we do not want to rely on the number
$n$ of nodes: this is cruical since at the end,
we'll apply the above algorithm for very small graphs with $m=o(n)$ edges.

So starting from the above algorithm $A_0$, we can build up a sequence
$A_k$ of algorithms, each having a better update cost till $k=\log m$
by the following lemma:

\begin{lemma}
\label{lem-akplusone}
    There is a universal constant $C$ such that if there exists an
    $(f(k)+g(k)\cdot m)$-ensuring algorithm $A_k$ for maximal matching,
    with a setup cost of $h(k,m,n)$,
    then there also exists an $(f(k)+C+\frac{g(k)}{2}\cdot m)$-ensuring
    algorithm $A_{k+1}$ for maximal matching as well, with a setup cost
    of $h(k,m,n)+O(n^2)$.
\end{lemma}

Before proving the above lemma, we derive the main result of the section. As $A_0$ is
an $c\cdot m$-ensuring algorithm, that is, $f(k)=0$ and $g(k)=c\cdot m$,
by induction we get the existence of an algorithm a $(k\cdot C+\frac{c}{2^k}\cdot m)$-ensuring algorithm for maximal matching. Now setting $k=\log m$ we get that
$A_{\log m}$ maintains a maximal matching with an amortized update cost
of $C\cdot \log m+\frac{2}{2^{\log m}}\cdot m=C\cdot\log m+2=O(\log m)$, thus
we get:

\begin{theorem}
	\label{thm-main-logmnegyzet}
	There exists a fully dynamic graph algorithm for maintaining a maximal
	matching with an $O(\log m)$ amortized update cost and constant query cost,
	using a lookahead of length $m$, with a setup cost of $O(n^2\cdot \log m)$.
\end{theorem}

Now we prove Lemma~\ref{lem-akplusone} by defining the algorithm $A_{k+1}$ below.
\begin{itemize}
	\item The algorithm $A_{k+1}$ works in \emph{phases} and returns a graph $G$ (as an edge list) and a matching $M$ (also as an edge list).
	\item The algorithm accesses the \emph{global} $\mate$ array in which the current maximal matching of the whole graph is stored.
		 ($A_{k+1}$ might get only a subgraph of the whole actual graph as input.)
	\item The algorithm manages a boolean array $T_{k+1}$ of size $n\times n$,
	initialized to be all-zero in the start of the program (hence the plus
	setup cost of $n^2$).
    \item As input, $A_{k+1}$ gets a graph $G$ and the update sequence $(q_1,\ldots,q_t)$, with a promise of $t\geq m_0$, where $m_0$ is the number of
    edges in $G$.
	\item The algorithm $A_{k+1}$ maintains a local matching $M$ as a list of
	edges (similarly to $A_0$), which is set to the empty list when calling $A_{k+1}$.
	\item In one phase, $A_{k+1}$ either handles a block
	$\vec{q}=(q_1,\ldots,q_{t'})$ of $t'$ operations for some $\frac{m_0}{4}\leq t'\leq \frac{m_0}{2}$, or a single operation.
	\item If $|G|$ is smaller than our favorite constant $42$, then the phase handles only the next update
	by explicitly modifying $G$, afterwards recomputing a maximal matching from scratch, in $O(42)$ (constant) time.
	That is,
	\begin{enumerate}
		\item We iterate through all the edges $(u,v)\in M$, and set $\textsc{mate}[u]$ and $\mate[v]$ to $\nul$ (in effect, we remove
		the ``local part'' $M$ of the global matching);
		\item We apply the next update operation on $G$;
		\item We set $M:=\textsc{Greedy}(G,\mate)$.
	\end{enumerate}
	\item Otherwise the phase handles $t'$ operations as follows. First, if
	there are at most $m_0$ unprocessed queries remaining (that can be checked
	by a lookahead of length $m_0\leq m$),
	then we finish the processing of the sequence
	in exactly two phases, each having $t'=\frac{t}{2}$ updates. Otherwise,
	we set $t'=\frac{m_0}{2}$, and handle the next $t'$ updates in a single phase.
	
	Observe that by this method, the value of $t'$ is always between $\frac{m_0}{4}$
	and $\frac{m_0}{2}$.
	
	\begin{enumerate}
	\item Using lookahead (observe that $t'<m$) we collect all the edges involved in $\vec{q}$ (either by an insert or a remove operation) into a graph $G'$.
	\item We iterate through all the edges $(u,v)\in M$, and set $\mate[u]:=\nul$, $\mate[v]:=\nul$.
	\item Iterating over all the edges $(u,v)$ in $G'$, we set $T_{k+1}(u,v)$
	and $T_{k+1}(u,v)$ to $1$.
	\item Using $T_{k+1}$ containing the adjacency matrix of $G'$ now,
	we split the list $G$ into the lists $G-G'$ and
	$G\cap G'$ by iterating through $G$ and putting $(u,v)$ to either $G-G'$
	(if $T_{k+1}(u,v)$ is zero) or to $G\cap G'$ (otherwise).
	\item We reset $T_{k+1}$ to be an all-zero matrix by iterating over $G'$
	again and resetting the corresponding entries.
	\item We run $M:=\textsc{greedy}(G-G',\mate)$.
	\item We call $A_k(G\cap G',(q_1,\ldots,q_{t'}))$. Let $G^*$ and $M^*$ be the graph and matching returned by $A_k$.
	\item We set $G:=(G-G')\cup G^*$ and $M:=M\cup M^*$.
	\end{enumerate}
\end{itemize}

In order to give the reader a better insight, we give an example before analyzing the time complexity.
To make the example more manageable, we adjust the constants as follows: we shall use the constant $1$ instead of $42$
(that is, if $G$ contains at most one edge, we do not make a recursive call but recompute the matching)
and also, the block size $A_2$ handles in one phase will be set to $1$ while $A_3$, which we call at the topmost level,
will handle $3$ operations in one phase.
\begin{example}
	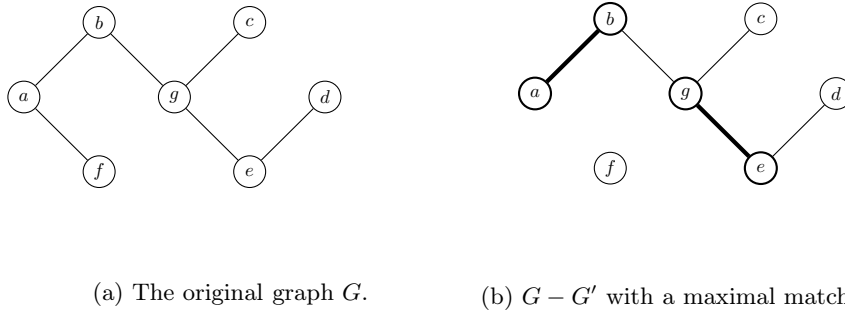
\begin{figure}[H]
		\centering\begin{subfigure}{13cc}
			\begin{adjustbox}{raise=1cm}
			\centering\begin{tikzpicture}[auto,node distance=2 cm, scale = 0.7, transform shape]
			
			\node[csucs] (a)  {$a$};
			\node[csucs] (b)  [above right of=a]  {$b$};
			\node[csucs] (f)  [below right of=a]  {$f$};
			\node[csucs] (g)  [above right of=f]  {$g$};
			\node[csucs] (c)  [above right of=g]  {$c$};
			\node[csucs] (d)  [below right of=c]  {$d$};
			\node[csucs] (e)  [below right of=g]  {$e$};
			
			
			\path[-] 
			(a) edge (b)
			(a) edge (f)
			(b) edge (g)
			(c) edge (g)
			(e) edge (g)
			(d) edge (e)
			;
			\end{tikzpicture}\end{adjustbox}
			\subcaption{The original graph $G$.}
		\end{subfigure}
		\begin{subfigure}{13cc}
			\centering\begin{adjustbox}{raise=1.1cm}
				\begin{tikzpicture}[auto,node distance=2 cm, scale = 0.7, transform shape]
				\node[csucs,thick] (a)  {$a$};
				\node[csucs,thick] (b)  [above right of=a]  {$b$};
				\node[csucs] (f)  [below right of=a]  {$f$};
				\node[csucs,thick] (g)  [above right of=f]  {$g$};
				\node[csucs] (c)  [above right of=g]  {$c$};
				\node[csucs] (d)  [below right of=c]  {$d$};
				\node[csucs,thick] (e)  [below right of=g]  {$e$};

				\path[-] 
				(a) edge [ultra thick] (b)
				(b) edge (g)
				(c) edge (g)
				(e) edge [ultra thick] (g)
				(d) edge (e)
				;
				\end{tikzpicture}\end{adjustbox}
			\subcaption{$G-G'$ with a maximal matching.}
		\end{subfigure}
		\hfil
	\caption{Executing Steps $1-3$ of $A_3$ on $G$, looking ahead the operations $+(f,g)$, $-(a,f)$, $+(d,c)$}
	\label{fig-matching-example-1}
	\end{figure}
	Let us assume that we call the algorithm $A_3$ on the graph $G=[(a,b), (b,g), (a,f),(g,e),(c,g),(d,e)]$ of Figure~\ref{fig-matching-example-1} (a). As the graph contains $6$ edges,
	which is more than our threshold $1$,
	a block of update operations of length $\frac{6}{2}=3$ will be handled in a phase, using lookahead.
	
	Now assume the next three update operations
	are $+(f,g)$, $-(a,f)$ and $+(d,c)$. Thus $G'=[(f,g),(a,f),(d,c)]$ is the list of edges involved, that's for Step 1. In Steps 2 and 3, we construct
	the graph $G''=G-G'$ and run the greedy matching algorithm on it, the (possible) result is shown in Figure~\ref{fig-matching-example-1} (b).
	(Note that the actual result depends on the order in which the
	edges are present in $G$.)
	
	In the Figure, thick circles denote those vertices having a non-$\nul$ mate at this point (that is, $\mate[a]=b$, $\mate[b]=a$, and so on, $c$, $d$ and $f$
	having a $\nul$ mate). Now, $A_2$ is called on $G\cap G'$ (depicted in Figure~\ref{fig-matching-example-2} (a)), and the whole block of three updates
	is passed to $A_2$ as well.
	
	\begin{figure}[h]
	\centering\begin{subfigure}{12cc}
		\centering\begin{tikzpicture}[auto,node distance=2 cm, scale = 0.7, transform shape]
		
		\node[csucs,ultra thick] (a)  {$a$};
		\node[csucs,ultra thick] (b)  [above right of=a]  {$b$};
		\node[csucs] (f)  [below right of=a]  {$f$};
		\node[csucs,ultra thick] (g)  [above right of=f]  {$g$};
		\node[csucs] (c)  [above right of=g]  {$c$};
		\node[csucs] (d)  [below right of=c]  {$d$};
		\node[csucs,ultra thick] (e)  [below right of=g]  {$e$};
		
		\path[-] 
		(a) edge (f)
		;
		\end{tikzpicture}
		\subcaption{The graph $G\cap G'$}		
	\end{subfigure}

	\centering\begin{subfigure}{12cc}
	\centering\begin{tikzpicture}[auto,node distance=2 cm, scale = 0.7, transform shape]
	
	\node[csucs,ultra thick] (a)  {$a$};
	\node[csucs,ultra thick] (b)  [above right of=a]  {$b$};
	\node[csucs] (f)  [below right of=a]  {$f$};
	\node[csucs,ultra thick] (g)  [above right of=f]  {$g$};
	\node[csucs] (c)  [above right of=g]  {$c$};
	\node[csucs] (d)  [below right of=c]  {$d$};
	\node[csucs,ultra thick] (e)  [below right of=g]  {$e$};
	
	\path[-] 
	(a) edge (f)
	(g) edge (f)
	;
	\end{tikzpicture}
	\subcaption{$A_2$ adds $(f,g)$ directly}
\end{subfigure}		\hfil
	\caption{Handling the first recursive call.}
	\label{fig-matching-example-2}
	\end{figure}
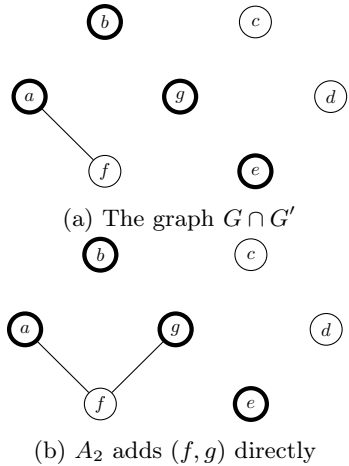
	Now as the input graph of $A_2$ has only one edge, $A_2$ just handles the next update $+(f,g)$;
	that is, it inserts the edge $(f,g)$ into its input of Figure~\ref{fig-matching-example-2} (a)
	and runs $\textsc{greedy}$ on this,
	resulting in the graph of Figure~\ref{fig-matching-example-2} (b).
	
	Observe that at this point $\textsc{mate}[a]=b$ and $\mate[g]=e$, so neither of these two edges is added to the maximal matching managed by $A_2$.
	That is due to the fact that the $\mate$ array is a global variable.
	This is vital: this way one can ensure that the union of the matchings of different recursion levels is still a matching, and also ensures a constant-time query cost.
	
	Then, as the current graph has two edges (which is larger than the threshold), $A_2$ handles a complete block of operations in a phase. (Now the length of the block happens to be $\frac{2}{2}=1$ so this
	does not make that much of a difference.)
	Thus, using a lookahead of length $1$, the only operation to be handled is $-(a,f)$. So we compute the difference graph and run $\textsc{greedy}$ on it (Figure~\ref{fig-matching-example-3} (a)),
	compute the intersection graph and call $A_1$ on this along with the update sequence consisting of the single operation $-(a,f)$ (Figure~\ref{fig-matching-example-3} (b)).
	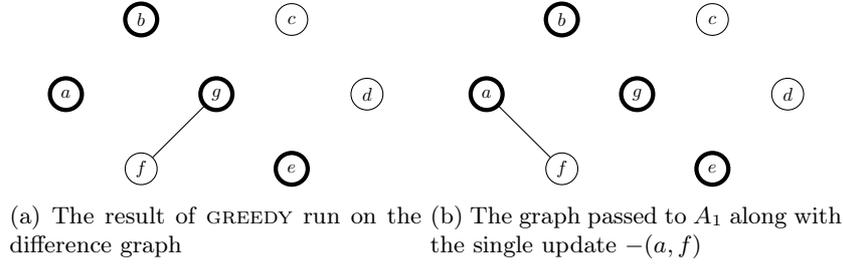
\begin{figure}[H]
		\centering\begin{subfigure}{12cc}
			\centering\begin{tikzpicture}[auto,node distance=2 cm, scale = 0.7, transform shape]
			
			\node[csucs,ultra thick] (a)  {$a$};
			\node[csucs,ultra thick] (b)  [above right of=a]  {$b$};
			\node[csucs] (f)  [below right of=a]  {$f$};
			\node[csucs,ultra thick] (g)  [above right of=f]  {$g$};
			\node[csucs] (c)  [above right of=g]  {$c$};
			\node[csucs] (d)  [below right of=c]  {$d$};
			\node[csucs,ultra thick] (e)  [below right of=g]  {$e$};
			
			\path[-] 
			(g) edge (f)
			;
			\end{tikzpicture}
			\subcaption{The result of $\textsc{greedy}$ run on the difference graph}
		\end{subfigure}
		\centering\begin{subfigure}{12cc}
			\centering\begin{tikzpicture}[auto,node distance=2 cm, scale = 0.7, transform shape]
			
			\node[csucs,ultra thick] (a)  {$a$};
			\node[csucs,ultra thick] (b)  [above right of=a]  {$b$};
			\node[csucs] (f)  [below right of=a]  {$f$};
			\node[csucs,ultra thick] (g)  [above right of=f]  {$g$};
			\node[csucs] (c)  [above right of=g]  {$c$};
			\node[csucs] (d)  [below right of=c]  {$d$};
			\node[csucs,ultra thick] (e)  [below right of=g]  {$e$};
			
			\path[-] 
			(a) edge (f)
			;
			\end{tikzpicture}
			\subcaption{The graph passed to $A_1$ along with the single update $-(a,f)$}
		\end{subfigure}
		\hfil
		\caption{Handling the second update}
		\label{fig-matching-example-3}
	\end{figure}
	As the input of $A_1$ is now a graph consisting of a single edge, it gets removed
	(as the edge in question is not involved in the matching, which can be seen e.g. from the $\mate$ array,
	the global matching is not changed), resulting in an empty graph on which $\textsc{greedy}$ gives
	an empty matching as well. Then, $A_1$ returns, as it handled the only operation it received.
	Now $A_2$ takes control. Concluding the second phase, it constructs the union of its intersection graph
	and the empty graph returned by $A_1$, so its current graph $G$ becomes the graph on Figure~\ref{fig-matching-example-3} (a).
	As now the graph has only one edge, the next update $+(d,c)$ is handled directly: the edge $(c,d)$ is inserted and $\textsc{greedy}$
	is run (Figure~\ref{fig-matching-example-4} (a)).
	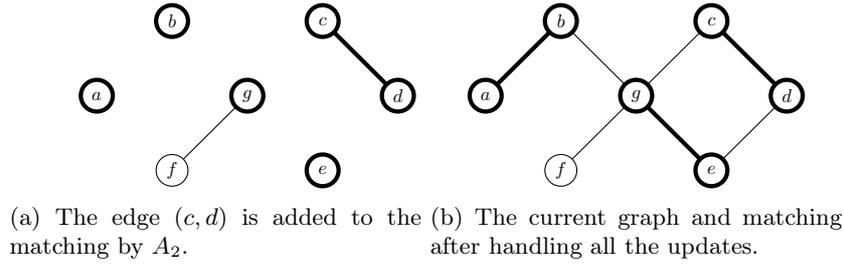
\begin{figure}[H]		
		\centering\begin{subfigure}{12cc}
		\centering\begin{tikzpicture}[auto,node distance=2 cm, scale = 0.7, transform shape]
		
		\node[csucs,ultra thick] (a)  {$a$};
		\node (name) [above left of=a] {};
		\node[csucs,ultra thick] (b)  [above right of=a]  {$b$};
		\node[csucs] (f)  [below right of=a]  {$f$};
		\node[csucs,ultra thick] (g)  [above right of=f]  {$g$};
		\node[csucs,ultra thick] (c)  [above right of=g]  {$c$};
		\node[csucs,ultra thick] (d)  [below right of=c]  {$d$};
		\node[csucs,ultra thick] (e)  [below right of=g]  {$e$};
		
		\path[-] 
		(g) edge (f)
		(c) edge[ultra thick] (d)
		;
		\end{tikzpicture}
		\subcaption{The edge $(c,d)$ is added to the matching by $A_2$.}
	\end{subfigure}
	\centering
	\begin{subfigure}{12cc}
	\centering\begin{tikzpicture}[auto,node distance=2 cm, scale = 0.7, transform shape]
	\node[csucs, ultra thick] (a)  {$a$};
	\node[csucs,ultra thick] (b)  [above right of=a]  {$b$};
	\node[csucs] (f)  [below right of=a]  {$f$};
	\node[csucs,ultra thick] (g)  [above right of=f]  {$g$};
	\node[csucs,ultra thick] (c)  [above right of=g]  {$c$};
	\node[csucs,ultra thick] (d)  [below right of=c]  {$d$};
	\node[csucs,ultra thick] (e)  [below right of=g]  {$e$};

	\path[-] 
	(a) edge [ultra thick] (b)
	(b) edge (g)
	(c) edge (g)
	(c) edge [ultra thick] (d)
	(e) edge [ultra thick] (g)
	(d) edge (e)
	(g) edge (f)
	;
	\end{tikzpicture}
	\subcaption{The current graph and matching after handling all the updates.}
\end{subfigure}
\hfil
		\caption{Handling the last update}
		\label{fig-matching-example-4}
	\end{figure}
	Now as $A_2$ has handled its whole input block, it returns its current graph: $A_3$ takes control and glues together its difference graph
	from Figure~\ref{fig-matching-example-1} (b) and the returned graph~\ref{fig-matching-example-4} (a), resulting in the graph
	in Figure~\ref{fig-matching-example-4} (b) which would be the starting graph of further updates.
\end{example}

Having completed this example, we will now show its correctness. That is, we claim that each $A_k$ maintains a maximal matching among those
vertices having a $\nul$ $\mate$ when the algorithm is called. This is true for the greedy algorithm $A_0$. Now assuming $A_k$ satisfies our
claim, let us check $A_{k+1}$. When the graph is small, the algorithm throws away its locally stored matching $M$, resetting the $\mate$
array to its original value in the process (in fact, this is the only reason why we store the local matching at each recursion level:
the global matching state can be queried by accessing the $\mate$ array alone). Then we handle the update and run $\textsc{greedy}$, which
is known to compute a maximal matching on the subgraph of $G$ spanned by the vertices having a $\nul$ mate. So this case is clear.

For the second case, if a block of $t'$ operations involving the edges
of the edge list $G'$ is handled, then we split the graph into two,
namely into a difference graph $G-G'$ and an intersection
graph $G'':=G\cap G'$. By construction, when handling the block, the edges belonging to $G-G'$ do not get touched and they are present in the graph during the whole
phase.

Hence, at any time point, a maximal matching
of $G$ can be computed by starting from a maximal matching of $G-G'$ and then extending the matching by a maximal matching in the subgraph of $G''$
not covered by the matching of $G'$. Thus, if we compute a maximal matching
$M'$ in the subgraph of $G-G'$ spanned by the vertices having a $\nul$ $\mate$, updating the $\mate$ array accordingly (that is, calling $\textsc{greedy}$
on $G'$), and maintaining a maximal matching $M''$ over the vertices of $G''$ having a $\nul$ $\mate$ after that point (which is done by $A_k$,
by the induction hypothesis), we get that at any time $M'\cup M''$ is a maximal matching of $G$. Hence, the algorithm is correct.

Now we analyze the time complexity of $A_{k+1}$.
Upon calling $A_{k+1}$, we set the local matching $M$ to be the empty list,
in constant time. Then, a phase either handles a single operation
(if $|G|$ is bounded by a constant threshold),
or a batch of $t'$ operations.

If $|G|$ is below the threshold $42$, then so is $|M|$,
thus running $\textsc{Greedy}$ also takes a constant time.

Assume the phase handles $t'$ operations for some $t'$ between $\frac{m_0}{4}$ and
$\frac{m_0}{2}$. Then, collecting the first $t'$ updates into a list $G'$ of
edges (containing possibly duplicates) takes $O(m_0)$ time. Now constructing
the intersection and the difference graphs maintaining an $O(m_0)$ time can
be done by using the global
boolean matrix $T_k$ of size $n\times n$, which is initalized to an
all-zero matrix in the very beginning of the program (hence, an initialization
cost of $n^2$ is needed to do that), then, the algorithm $A_k$ sets
those entries $T_k(u,v)$ and $T_k(v,u)$ for which $(u,v)\in G'$ to one.
Using $T_k$, the list $G$ can be split into $G-G'$ and $G\cap G'$ using
$O(m_0)$ time. Then, as $G-G'$ also has at most $m_0$ edges, $\textsc{Greedy}$
runs in $O(m_0)$ steps on it as well.

Then, we call $A_k(G\cap G',(q_1,\ldots,q_{t'}))$. Observe that since $G'$
is the graph constructed from the $t'$ queries, it has at most $t'$ edges,
hence $|G\cap G'|\leq t'$. Now by assumption, $A_k$ guarantees in this case
that the queries can be processed in an amortized time of $f(k)+g(k)\cdot t'$,
since $t'$ is an upper bound for the size of this dynamic graph during
its whole lifecycle. As $\frac{m_0}{4}\leq t'\frac{m_0}{2}$, this gives
an amortized cost at most $f(k)+\frac{g(k)}{2}\cdot m_0$ per update.

Finally, at the end of the phase we have to concatenate the two lists
containing the graphs $G-G'$ and $G^*$ returned by $A_k$, and the two
matchings $M$ and $M^*$, and clear the entries $(u,v)$ of $T_k$ for which
$(u,v)$ is present in $G'$ (this is needed to ensure that at the beginning
of each phase, $T_k$ is an all-zero helper matrix). This can be done in
$O(m_0)$ steps as well, by simply iterating through $G'$.

Overall, to process the $t'$ updates, the algorithm takes an amortized cost
of $f(k)+\frac{g(k)}{2}\cdot m_0$ per update, plus a total cost of $O(m_0)$ 
for some universal constant $C$, which makes the amortized cost to be
$C+f(k)+\frac{g(k)}{2}\cdot m_0$ for some universal constant $C$,
since the number $t'$ of updates is at least $\frac{m_0}{4}$.

Since in each phase $m_0\leq m$ (as $m_0$ is the size of the graph $G$
in a specific time point, while $m$ is the maximum of those values over
time), we proved Lemma~\ref{lem-akplusone} and thus Theorem~\ref{thm-main-logmnegyzet}. \hfil$\Box$

\subsection{Implementation details and improving the setup cost}
The careful reader might observe the fact that the algorithms $A_k$ never use
the value $m$ to make decisions, neither in the length of the lookahead it
uses, nor for setting the length of $t'$. Hence the amortized update cost
is guaranteed to be an actual $O(\log m)$. Also, the sequence of these
algorithms can be constructed as a single algorithm $A$, taking as argument
a graph $G$, the sequence $\vec{q}$ of updates, and the recursion depth $k$
as an integer -- but this latter value is used only
in order to determine which helper table $T_{k}$ can $A$ use when constructing
the graphs $G-G'$ and $G\cap G'$. However, the construction of these graphs
happens during substeps $3-5$ in which no recursive call is made, and after
which $T_k$ is again guaranteed to be an all-zero matrix -- hence, the algorithm
$A$ can use the very same helper array $T$ on each recursion level.
This already improves the setup cost to be $O(n^2)$ instead of $O(n^2\cdot \log m)$
as there is only one adjacency matrix $n\times n$ we have to handle globally,
which we initialize by zeroes.

However, we can do even better: during the construction of $G-G'$ and $G\cap G'$,
we only check those entries of $T$ which correspond to edges already present
in $G$. Thus, we can postpone the initialization: it suffices to set $T(u,v)$
and $T(v,u)$ to zero only for those entries for which $(u,v)$ is in $G$,
which can be handled during the processing of an insert operation
for a constant increase in the amortized run-time.

The $\mate$ array has to be initialized to an all-zero vector in the beginning,
though, requiring an $O(n)$ setup cost.

So the final form of our main result is the following:
\begin{theorem}
	There exists a fully dynamic graph algorithm for maintaining a
	maximal matching with an $O(\log m)$ amortized update cost and
	constant query cost, using a lookahead of length $m$, with either
	an $O(n)$ setup cost (if the memory model allows getting an uninitialized
	memory chunk of size $n\times n$ in constant time)
	or a setup cost of $O(n^2)$ (if in the memory model we have to pay $n^2$
	even when the memory is uninitialized).
\end{theorem}

Note that in the latter case if $O(n^2)$ is too much of a cost for either
in storage space, or as a setup cost, then the set operations required
for splitting the graph $G$ can be implemented by using balanced binary trees
for the set operations. That way, splitting the list $G$ of size $m_0$,
we can make a searchable set from $G'$ in $O(t'\cdot \log t')$ time,
then doing a search operation for each element of the list $G$ takes
an additional time of $O(m_0\cdot \log t')$. As $t'\leq m_0$, that's a total
time of $O(m_0\log m_0)$ which yields and additional $\log m_0$ amortized cost
per update on each recursion level. This way, an amortized update cost
of $O(\log^2m)$ can be gained, for a setup cost of $O(n)$, that only uses
a single global $\mate$ array of size $n$, and lists/sets having in total $O(m)$
elements, with a setup cost of $O(n)$, which might be a more memory-efficient
solution for graphs which are guaranteed to be sparse at any given time.

If even maintaining the array $\mate$ is too much, then one can trade it
for a global tree map in which those vertices having a mate appear as key,
with their mate as value. That decision makes $\mate$ accesses to have the
cost of $O(\log m)$ (as there are at most $2m$ nodes actually having a mate
at each time step: bear in mind that $O(\log m)$ is automatically $O(\log n)$
as well, but not necessarily vice versa), allowing for a constant initialization
cost and a total memory needed is only that of storing $O(m)$ nodes/edges.
The query cost becomes $O(\log m)$ in that case. Managing the $\mate$ tree map
in the code yields an additional total cost of $O(m_0\times\log m)$
(erasing the local part of the matching) for a phase, which translates to
an additional amortized cost of $\log m$ per update -- which is free if we
already traded the helper array $T$ for set-operations.

So in that case we get an algorithm with an amortized update cost of $O(\log^2 m)$,
query cost of $O(\log m)$, a constant setup cost and a memory footprint proportional
to storing $O(m)$ nodes. This might be the correct choice if we do not know the
size of the graph in advance, or if the nodes are not numbers but strings, say,
whose possible domain is not known in advance.

\section{Conclusion}
In this study we dealt with a problem arising in the context of fully dynamic graph algorithms.
We showed that by using a \emph{lookahead} of linear length,
there is a \emph{deterministic} algorithm achieving an $O(\log m)$ amortized update cost, without knowing the maximal size $m$ of the graph in advance. (Note that
once again that $O(\log m)$ is $O(\log n)$ as well, since $m\leq n^2$.)

This result shows that lookahead can help in the dynamic setting for problems other than the
transitive closure (and the SCC) properties, studied in~\cite{lookahead}: indeed, the best known deterministic algorithm
for the problem using no lookahead has an update cost of $O(\sqrt{m})$.

It is an interesting question to study further the possibilities of using lookahead for different problems,
and maybe factor in also randomization as well, albeit for the randomized setting,
an algorithm with a (both expected and whp) constant update cost is already known
without lookahead.
{
\bibliography{biblio}{}
\bibliographystyle{plain}
}

\end{document}